\documentclass[reprint, amsmath,amssymb,aps]{revtex4-2}
\usepackage{svg}
\usepackage{graphicx}
\usepackage{dcolumn}
\usepackage{gensymb}
\usepackage{bm}
\usepackage{mathtools}
\usepackage[colorlinks=true, linkcolor=blue,anchorcolor=blue, citecolor=blue,urlcolor=blue]{hyperref}

\usepackage{amssymb, amsmath, amsthm}
\usepackage{booktabs}
\usepackage{siunitx}
\usepackage{multirow}
\usepackage{dcolumn}
\usepackage{hyperref}
\usepackage{enumerate}
\hypersetup{
	colorlinks=true,
	linkcolor=blue,
	filecolor=gray,
	urlcolor=blue,
	citecolor=blue,
}
\begin{document}
	
	\preprint{APS/123-QED}
	
	\title{Deep learning density functional theory Hamiltonian in real space}

	\author{Zilong \surname{Yuan}$^{1}$}
	\thanks{These authors contributed equally to this work.}
	
	\author{Zechen \surname{Tang}$^{1}$}
	\thanks{These authors contributed equally to this work.}
	
	\author{Honggeng \surname{Tao}$^{1}$}
	\thanks{These authors contributed equally to this work.}
	
	\author{Xiaoxun \surname{Gong}$^{1,2,3}$}
	\author{Zezhou \surname{Chen}$^{1}$}
	\author{Yuxiang \surname{Wang}$^{1}$}
	\author{He \surname{Li}$^{1,4}$}
	\author{Yang \surname{Li}$^{1}$}
	\author{Zhiming \surname{Xu}$^{1}$}
	\author{Minghui \surname{Sun}$^{1}$}
	\author{Boheng \surname{Zhao}$^{1}$}
	\author{Chong \surname{Wang}$^{1}$}
	
	\author{Wenhui \surname{Duan}$^{1,4,5}$}
	\email{duanw@tsinghua.edu.cn}
	\author{Yong \surname{Xu}$^{1,5,6}$}
	\email{yongxu@mail.tsinghua.edu.cn}
	\affiliation{$^{1}$State Key Laboratory of Low Dimensional Quantum Physics and Department of Physics, \\Tsinghua University, Beijing 100084, China}
	\affiliation{$^{2}$Department of Physics, University of California at Berkeley, California 94720, USA}
	\affiliation{$^{3}$Materials Sciences Division, Lawrence Berkeley National Laboratory, Berkeley, California 94720, USA}
	\affiliation{$^{4}$Institute for Advanced Study, Tsinghua University, Beijing 100084, China}
	
	\affiliation{$^{5}$Frontier Science Center for Quantum Information, Beijing 100084, China}
	\affiliation{$^{6}$RIKEN Center for Emergent Matter Science (CEMS), Wako, Saitama 351-0198, Japan}

	\begin{abstract}
		Deep learning electronic structures from ab initio calculations holds great potential to revolutionize computational materials studies. While existing methods proved success in deep-learning density functional theory (DFT) Hamiltonian matrices, they are limited to DFT programs using localized atomic-like bases and heavily depend on the form of the bases. Here, we propose the DeepH-r method for deep-learning DFT Hamiltonians in real space, facilitating the prediction of DFT Hamiltonian in a basis-independent manner. An equivariant neural network architecture for modeling the real-space DFT potential is developed, targeting a more fundamental quantity in DFT. The real-space potential exhibits simplified principles of equivariance and enhanced nearsightedness, further boosting the performance of deep learning. When applied to evaluate the Hamiltonian matrix, this method significantly improved in accuracy, as exemplified in multiple case studies. Given the abundance of data in the real-space potential, this work may pave a novel pathway for establishing a ``large materials model" with increased accuracy.
	\end{abstract}
	\maketitle

	\section{Introduction}
	The integration of deep-learning methods with density functional theory (DFT) holds great promise for accelerating computational materials simulations at a first-principles level, providing possible solutions for the ``accuracy-efficiency dilemma" of DFT~\cite{behler_2007,schutt_schnet_2017,zhang_deep_2018,Unke2021,Gu2022,batzner_e3-equivariant_2022,li2022deep,gong_general_2023,tang2023efficient,li2023deep,li2024deep,wang2024deeph,WANG2024,li2024neural,yuan_equivariant_2024,tang2024densitymatrix,yu2023efficient,phisnet_2021,liao2023equiformerv2}. Within this emerging research field, deep-learning Hamiltonian (DeepH) methods are gaining increasing interest~\cite{li2022deep,gong_general_2023,wang2024deeph}. The DFT Hamiltonian serves as a fundamental quantity in DFT computations, from which all physical quantities at single-particle level can be derived. Moreover, the DFT Hamiltonian bears fundamental physical priors of equivariance and nearsightedness~\cite{Kohn1996,Prodan2005}, which have been effectively levitated in DeepH to enhance performance. Former research has achieved meV-level accuracy in Hamiltonian matrix elements~\cite{gong_general_2023}, demonstrating both precise modeling and exceptional transferability~\cite{WANG2024}. Despite general successes, previous research on the DFT Hamiltonians $\hat{H}_{\text{DFT}}$ has predominantly focused on modeling the Hamiltonian matrix elements under localized atomic-like bases, represented as $H_{\alpha\beta}$, with $\alpha$ and $\beta$ denoting basis indices. Although carefully constructed basis sets facilitate the efficient representation of the electronic structure in a low-dimensional space, the resulting Hamiltonian matrix heavily depends on the form and reliability of the basis. In this sense, $H_{\alpha\beta}$ provides a less ``fundamental" representation of $\hat{H}_{\text{DFT}}$ due to its basis-dependent nature. Additionally, previous frameworks for deep-learning Hamiltonians are generally incompatible with DFT programs with plane-wave bases, limiting their application on numerous DFT programs, particularly in solid systems. 
	
	A potential solution to these issues lies in developing a deep-learning Hamiltonian framework that models a basis-independent quantity. This is feasible through the formulation of the Kohn-Sham DFT, where the electronic Hamiltonian consists of the electronic kinetic operator $\hat{T}_s$ and a local Kohn-Sham potential $\hat{V}_{\text{KS}}(\textbf{r})$. Since $\hat{T}_s$ is independent from DFT self-consistent iterations and can be computed at negligible cost, $\hat{V}_{\text{KS}}(\textbf{r})$ contains sufficient information to reproduce $\hat{H}_{\text{DFT}}$~\cite{Martin2004}. Furthermore, $\hat{V}_{\text{KS}}(\textbf{r})$ is a basis-independent quantity and can be accessed from DFT codes employing any types of basis sets.
	
	\begin{figure*}[htbp]
		\centering
		\includegraphics[width=0.9\linewidth]{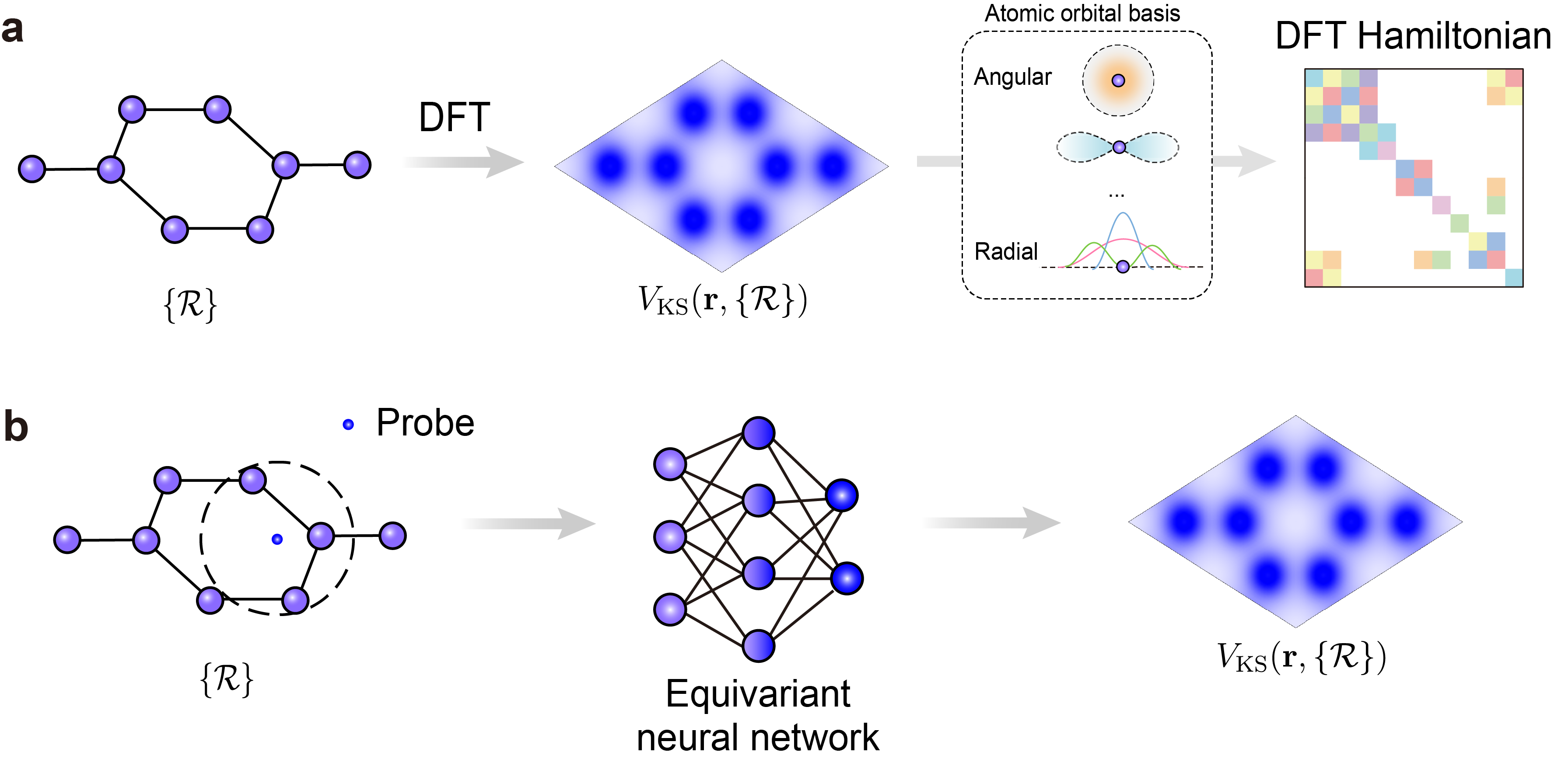}
		\hspace{0.5cm}
		\caption{Schematic illustration of DeepH-r. (a) The real-space Kohn-Sham potential can be obtained by conventional DFT calculations, and integrated under certain bases into the DFT Hamiltonian matrix. (b) Material structure is mapped into neural network features, and the real-space Kohn-Sham potential is predicted by adding probe nodes at spatial points in the DeepH-r method. The predicted $V_{\text{KS}}$ can then be utilized in further computations in the same manner as in (a) for subsequent computations.} 
		\label{fig:1}
	\end{figure*}
	
	In this work, we developed an equivariant neural-network framework termed ``DeepH-r" (Deep-learning Hamiltonian in real space) to realize the deep-learning modeling of $\hat{V}_{\text{KS}}(\textbf{r})$. This task involves establishing a mapping from the atomic structure $\{\mathcal{R}\}$ to the Kohn-Sham potential $\hat{V}_{\text{KS}}(\textbf{r},\{\mathcal{R}\})$. In contrast to conventional DFT, which establishes this mapping through self-consistent iterations (Fig.~\ref{fig:1}a), DeepH-r encodes the mapping from atomic structures to real-space quantities using an equivariant neural network. 
	
	To represent $\hat{V}_{\text{KS}}$, a ``probe node" can be added at $\textbf{r}$ to predict $\hat{V}_{\text{KS}}(\textbf{r},\{\mathcal{R}\})$, as demonstrated in Fig.~\ref{fig:1}b.
	The predicted Kohn-Sham potential can then be utilized to derive the DFT Hamiltonian under different basis sets via spatial integration, following the same workflow as conventional DFT computations(Fig.~\ref{fig:1}a). 
	This approach entirely bypasses the self-consistent iterations, which are the most time-consuming steps in the Kohn-Sham DFT framework.
	
	Upon closer investigation of deep-learning $\hat{V}_{\text{KS}}$, we identified at least three major benefits of this quantity, in addition to its basis-independent nature: 1. $\hat{V}_{\text{KS}}$ encompasses a simplified form of equivariance, enabling more flexible neural-network design of DeepH-r; 2. As a quantity defined on real-space points, $\hat{V}_{\text{KS}}$ exhibits enhanced nearsightedness compared with the Hamiltonian matrix, even when using localized bases; 3. $\hat{V}_{\text{KS}}$ provides more comprehensive data of DFT computations than the Hamiltonian, and the increased dataset size is often favorable for deep-learning. Further discussions of the enhanced physical priors are available in the ``Methods" section. These enhanced physical priors of $\hat{V}_{\text{KS}}$ leads to significantly improved performance of DeepH-r, as exemplified by multiple case studies shown in the ``Results" section.
	
	\section{Methods}
	
	\subsection{Physical priors of $\hat{V}_{\text{KS}}$}
	Two physical priors of the Hamiltonian, equivariance and nearsightedness, play pivotal roles in the deep learning of DFT Hamiltonian. Leveraging these priors is beneficial for neural network performance, and functions as the designing principles for predicting the Hamiltonian~\cite{li2022deep}. Through an investigation of $\hat{V}_{\text{KS}}$, we find that these physical priors not only hold, but are also either simplified or strengthened compared with the Hamiltonian matrix $\hat{H}_{\alpha\beta}$. This provides the potential to enhance the performance in predicting Hamiltonian via DeepH-r.
	
	From the equivariance perspective, the mapping of material structure to $V_{\text{KS}}{(\mathbf{r},\{\mathcal R\})}$ is also equivariant with respect to the $\text{E}(3)$ group (the Euclidean group in three-dimensional space), comprising spatial translation, $\text{SO}(3)$ rotation, and spatial inversion. Notably, the equivariance of $V_{\text{KS}}{(\mathbf{r},\{\mathcal R\})}$ with respect to rotation $\hat{\textbf{R}}$ is greatly simplified, taking the form $V_{\text{KS}}{(\mathbf{r},\{\mathcal R\})} =V_{\text{KS}}(\hat{\mathbf{R}} \cdot \mathbf{r},\hat{\mathbf{R}} \cdot \{\mathcal R\}) $. In this sense, $\hat{V_{\text{KS}}}$ is invariant under rotation, provided the spatial points rotate simultaneously with the material structure. In contrast, the Hamiltonian matrix elements under atomic bases ($H_{\alpha\beta}$) is $\text{SO}(3)$ equivariant, carrying a representation of order $l_{\alpha}\otimes l_{\beta}$ of the $\text{SO}(3)$ group, where $l_\alpha$ and $l_\beta$ denotes angular momentum of bases $\alpha$ and $\beta$, respectively. The higher angular momentum introduced by the bases brings additional complexity to the neural network design. Meanwhile, the angular dependence of physical properties in solid materials may be described with a lower order of $l$, and the higher angular momentum may be unnecessary for representing the electronic structure. Predicting $\hat{V}_{\text{KS}}$, in contrast, eliminates such artificial complexity.
	
	$\hat{V}_{\text{KS}}$ exhibits enhanced nearsightedness compared to $H_{\alpha\beta}$. Specifically, $H_{\alpha\beta}$ is computed as a sum of kinetic term and the integration $\int d\mathbf{r}\phi_{\alpha}^*(\mathbf{r})\hat{V}_{\text{KS}}(\mathbf{r})\phi_{\beta}(\mathbf{r})$. The nearsightedness of $H_{\alpha\beta}$ is present if $\phi_\alpha$ and $\phi_\beta$ are localized bases, but the nearsightedness is indirect. Atomic bases have a spatial spread ranging from $5$ to $10$~bohr, making $H_{\alpha\beta}$ sensitive to any changes of $\hat{V}_{\text{KS}}(\mathbf{r})$ within this range, thereby weakens its nearsightedness. In contrast, the basis-free $\hat{V}_{\text{KS}}(\mathbf{r})$, defined on real-space points, bears an enhanced form of nearsightedness. Such feature is beneficial for both the neural-network design and the generalization capability of DeepH-r.
	
	\subsection{Preconditioning of $\hat{V}_{\text{KS}}$}
	
	Followed by the investigation of the physical priors of $\hat{V}_{\text{KS}}$, another critical issue is determining which parts of the $\hat{V}_{\text{KS}}(\textbf{r})$ should be predicted to efficiently reproduce electronic structure properties, including $\hat{H}_{\text{DFT}}$. Within the Kohn-Sham DFT framework, the Hamiltonian is described as:
	
	\begin{equation}
		\label{H_DFT}
		\hat{H}_{\text{DFT}}[n] = \hat{T}_s + \hat{V}_{\text{ext}} + \hat{V}_{\text{Hartree}}[n](\textbf{r}) + \hat{V}_{\text{xc}}[n](\textbf{r}) , \nonumber
	\end{equation}
	
	where the four terms represent the single-electron kinetic operator, the external potential from the atomic nuclei, the classical Coulomb repulsion between electrons (the Hartree term), and the exchange-correlation potential. The sum of the latter three terms is regarded as Kohn-Sham potential $\hat{V}_{\text{KS}}$. $\hat{T}_s$ is independent of self-consistent iteration and can be evaluated at negligible cost via analytical formula or two-center integration~\cite{soler2002siesta}. $\hat{V}_{\text{ext}}(\textbf{r})$ depends on the pseudopotential applied. For instance, within the norm-conserving pseudopotential framework~\cite{ncpp1979,kbproj1982}, the external potential comprises a non-local term $\hat{V}_{\text{KB}}(\textbf{r},\textbf{r}')$ and a Coulomb-like local term $\hat{V}_{\text{ion}}(\textbf{r})$. The non-local term could be efficiently computed without self-consistent iterations, and is not treated in DeepH-r. 
	
	Regarding the Coulomb-like potentials including $\hat{V}_{\text{ion}}$ and $\hat{V}_{\text{Hartree}}$, a ``screening" technique is introduced by the SIESTA program to further split these terms~\cite{soler2002siesta}. Specifically, $\hat{V}_{\text{Hartree}}$ is split into a self-consistent-independent part and a part dependent on the self-consistent iterations. The former term is combined with $\hat{V}_{\text{ion}}$ to form a screened ``neutral atom" potential $\hat{V}_{\text{NA}}$. This neutral atom potential is a superposition of atom-centered localized potential of each atom, independent of self-consistent iterations, and could be computed at negligible cost. The remaining term in $\hat{V}_{\text{Hartree}}$ is denoted $\delta \hat{V}_{\text{Hartree}}$, corresponding to the Coulomb potential generated by a charge density which is smaller in value, and exhibits a smoother real-space distribution~\cite{soler2002siesta}.
	
	Consequently, DeepH-r can focus on predicting only $\delta \hat{V}_{\text{Hartree}}(\textbf{r})+\hat{V}_{\text{xc}}(\textbf{r})$. After prediction, $\hat{T}_s, \hat{V}_{\text{KB}}$ and $\hat{V}_{\text{NA}}$ are efficiently evaluated through non-self-consistent computation. Summing these terms provides the full $\hat{H}_{\text{DFT}}$, facilitating the computation of other electronic structure quantities. This strategy eliminates$\hat{V}_{\text{NA}}(\mathbf{r})$ which changes dramatically near atomic nuclei, leaving only a smooth potential for DeepH-r to predict. As exemplified in ablation studies~\cite{supp}, this strategy leads to significant improvement in accuracy, compared to predicting the less smooth total potential. While the previous discussions have taken norm-conserving pseudopotential and localized basis DFT as an example, such preconditioning strategy could be easily to generalized into other methods, including all-electron computations or various other pseudopotentials, and DFT programs utilizing plane-wave or finite element bases.

	\subsection{Neural network architecture}
	DeepH-r incorporates a SE(3)-equivariant neural network architecture. When mapping atomic structures into neural network features, only relative atomic positions are considered, ensuring the translation equivariance of DeepH-r. For the $\text{SO(3)}$ rotational equivariance, each network feature (including the output features) is an equivariant vector carrying a specific irreducible representation of the $\text{SO(3)}$ group of order $l$. Upon spatial rotation $\hat{\textbf{R}}$ of the material structure, each equivariant vector $\textbf{x}^l$ transforms to $D^l({\textbf{R}})\textbf{x}^l$, where $D^l(\textbf{R})$ denoting the Wigner-D matrix, thereby guaranteeing the $\text{SO(3)}$ equivariance of DeepH-r. All neural network operations are carefully designed to maintain the equivariance, among which the equivariant tensor product is the most time-consuming operation. To enhance efficiency, DeepH-r employs a more efficient tensor product based on $\text{SO}(2)$ equivariance under local coordinates, as proposed by eSCN~\cite{passaro2023reducing}.
	
	The building blocks of DeepH-r are message-passing blocks between nearby features. Along with the eSCN tensor product, equivariant attention mechanisms are also introduced in these blocks, in analogy with related works~\cite{liao2023equiformerv2,wang2024deeph}. For the prediction of real-space $\hat{V}_{\text{KS}}$, two types of message-passing blocks are introduced, including the atomic blocks and the probe blocks. 
	The atomic blocks aim to update atomic features with their surrounding environment. 
	The probe blocks are designed to aggregate the nearby learned atomic features into the probe features and predicting $\hat{V}_{\text{KS}}$ at corresponding probe coordinates.
	
	Integrating advanced neural-network techniques, DeepH-r represents the first attempt to predict Hamiltonian via real space.
	
	\begin{figure*}[htp]
		\centering
		\includegraphics[width=0.7\linewidth]{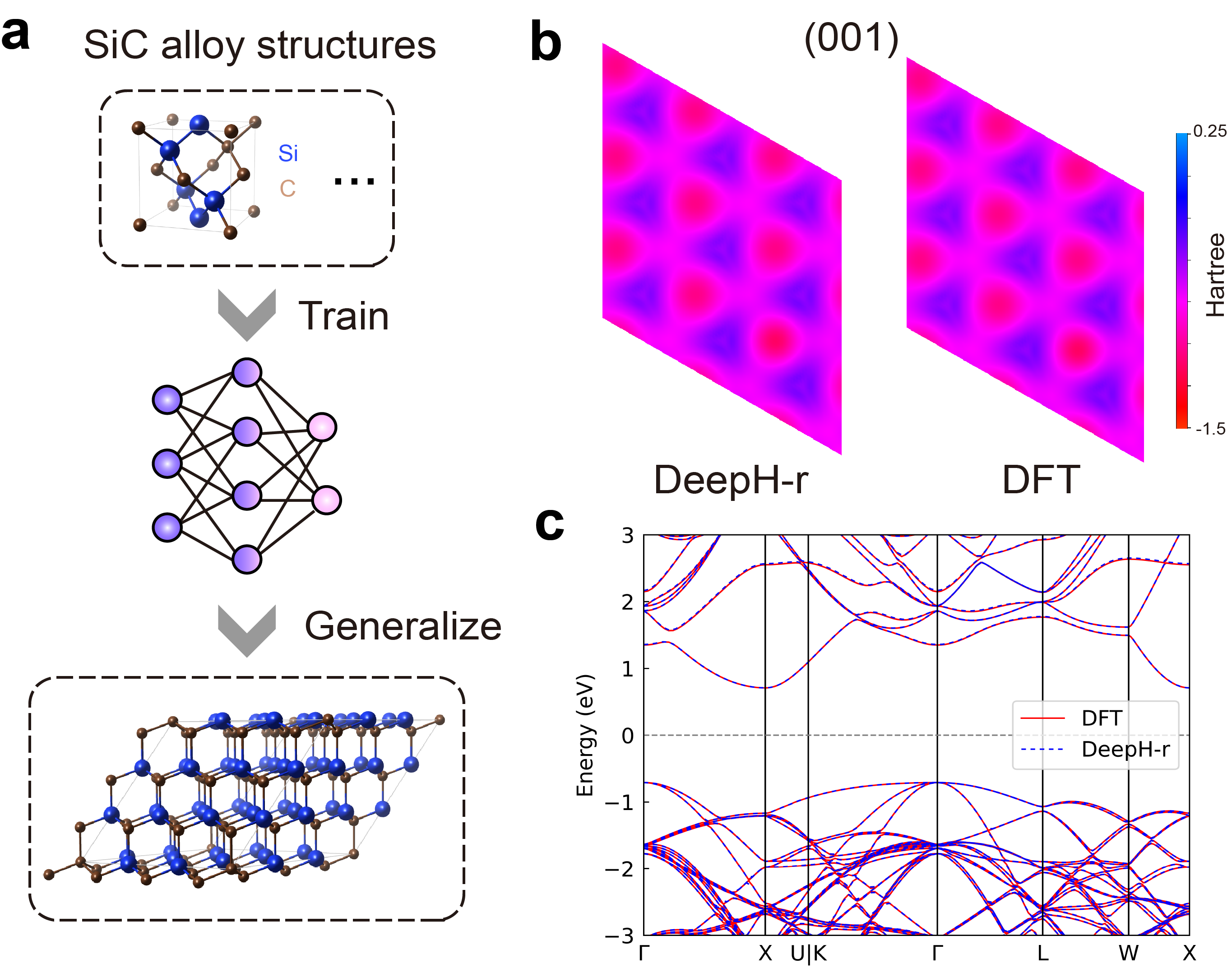}
		\hspace{0.5cm}
		\caption{Application of the DeepH-r on the SiC alloy system. (a) DeepH-r is trained on small-size SiC alloy structures with random substitutions between Si and C, and generalized to the  3 × 3 × 3 SiC supercell with a single Si atom substituted by C. (b,c) Comparison of DFT calculated and DeepH-r predicted (b) real-space distributions of the $V(\textbf{r})$ within the (001) plane and (c) the electronic band structure of the 3 × 3 × 3 SiC structure with a single Si atom substituted by C.}   
		\label{fig:3}
	\end{figure*}
	
	\section{Results}
	
	We first benchmark the performance of DeepH-r by the DeepH-E3 method\cite{gong_general_2023}  on the  perturbed $\beta$-SiC and SiC alloy datsets. DeepH-r and DeepH-E3 are trained under the same data splitting of training, validation and test sets with the ratio of  6 : 2 : 2. 
	Our experiments show that the mean absolute errors (MAEs) of  
	DFT Hamiltonian matrix elements from the DeepH-r significantly surpasses the DeepH-E3 in both datasets, demonstrating the remarkable improvements and high accuracy of the DeepH-r method. Furthermore, we employ the DeepH-r trained on small SiC alloy structures with random substitutions to generalize on the large SiC pristine structure with single C atom substituted as shown in Fig.~\ref{fig:3} a. As shown in Fig.~\ref{fig:3} b and Fig.~\ref{fig:3} c, the real space distribution of $V(\textbf{r})$ in the (001) plane and the band structure predicted by the DeepH-r are in good agreement with the DFT benchmark data. Notably, the training set generation process enforces fully random substitutions, containing only training structures with a dense defect concentration (around $50\%$). Despite being trained on such dataset, DeepH-r still exhibits good generalization to structures with a single defect (corresponding to a defect concentration of only $4\%$), demonstrating its generalization ability. Details are described in the supplementary material \cite{supp}. 
	
	\begin{figure*}[htbp]
		\centering
		\includegraphics[width=1\linewidth]{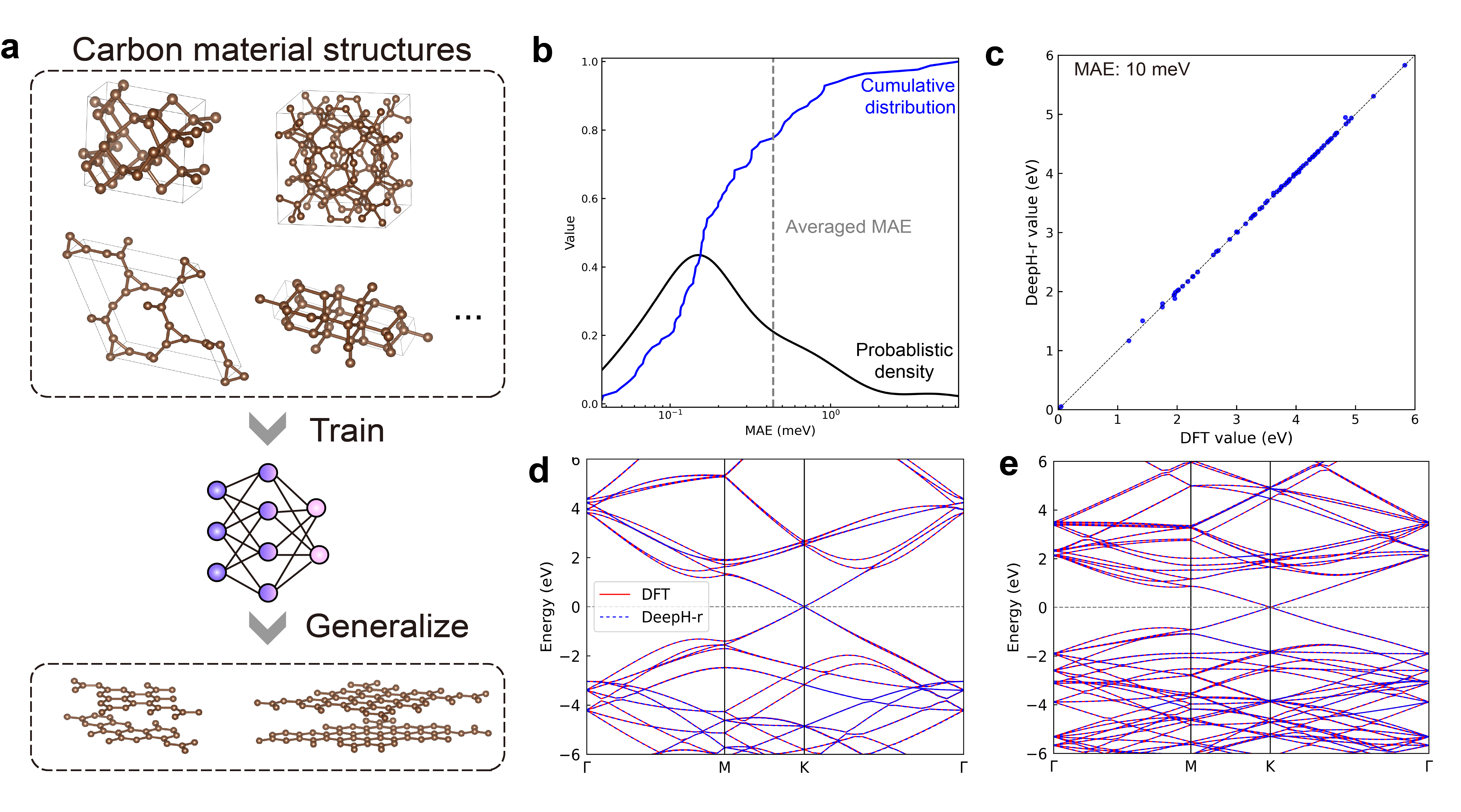}
		\hspace{0.5cm}
		\caption{Application of DeepH-r on the carbon allotropes dataset.(a) DeepH-r is trained on various carbon allotropes and generalized on the (2,1) and (3,2) twisted bilayer graphenes (TBGs). (b) Cumulative distribution and probability density  across test structures, giving an overall MAE of 0.44 meV. (c) Comparisons of band gap calculated by DFT and predicted by DeepH-r on the test set of carbon allotropes. (d,e) Comparison of DFT calculated and DeepH-r predicted band structures of the (2,1) and (3,2) TBGs.}
		\label{fig:4}
	\end{figure*}
	
	As a more challenging case study, the DeepH-r method is employed to study carbon allotropes with diverse atomic structures. The carbon materials dataset is sourced from the Samara Carbon Allotrope Database (SACADA)~\cite{SACADA2016}, comprising of 427 carbon allotrope structures. Significantly, the MAE of DFT Hamiltonian matrix predicted by DeepH-r on the test set is as low as 0.44 meV as shown in 
	Fig.~\ref{fig:4} b.
	This ensures the  reliable and accurate predictions of band gaps, giving MAEs of 10 meV as shown in  Fig.~\ref{fig:4} c. We further use the trained DeepH-r model to study twisted bilayer graphenes (TBGs). As shown in Fig.~\ref{fig:4} (d,e), the band structures of (2,1) and (3,2) TBGs are well 
	consistent with the DFT benchmark data. While only bulk carbon alloytropes are present in the dataset, DeepH-r still bears transerbility to these low-dimensional systems, fully demonstrating its high precision and good capability to generalize.  
	
	\section{Conclusions}
	In summary, we introduce DeepH-r, a deep learning framework for predicting the real space Kohn-Sham potential. 
	Benefiting from the enhanced physical priors and the basis-independent nature of the real space potential, the DeepH-r method has demonstrated the capability to accurately capture this fundamental DFT quantity, as exemplified in the case studies of SiC and carbon allotropes.
	Our work not only opens a new pathway for deep-learning electronic structure, but also makes significant improvements on the DeepH approaches, opening the new opportunities for high-accuracy and universal electronic structure calculations.

	\begin{acknowledgments}
		This work was supported by the Basic Science Center Project of NSFC (grant no. 52388201),  the National Key Basic Research and Development Program of China (grant no. 2023YFA1406400), the National Natural Science Foundation of China (grant no. 12334003), the National Science Fund for Distinguished Young Scholars (grant no. 12025405), the Beijing Advanced Innovation Center for Future Chip (ICFC), and the Beijing Advanced Innovation Center for Materials Genome Engineering. The work was carried out at National Supercomputer Center in Tianjin using the Tianhe new generation supercomputer.

	\end{acknowledgments}


\begin{thebibliography}{28}%
	\makeatletter
	\providecommand \@ifxundefined [1]{%
	 \@ifx{#1\undefined}
	}%
	\providecommand \@ifnum [1]{%
	 \ifnum #1\expandafter \@firstoftwo
	 \else \expandafter \@secondoftwo
	 \fi
	}%
	\providecommand \@ifx [1]{%
	 \ifx #1\expandafter \@firstoftwo
	 \else \expandafter \@secondoftwo
	 \fi
	}%
	\providecommand \natexlab [1]{#1}%
	\providecommand \enquote  [1]{``#1''}%
	\providecommand \bibnamefont  [1]{#1}%
	\providecommand \bibfnamefont [1]{#1}%
	\providecommand \citenamefont [1]{#1}%
	\providecommand \href@noop [0]{\@secondoftwo}%
	\providecommand \href [0]{\begingroup \@sanitize@url \@href}%
	\providecommand \@href[1]{\@@startlink{#1}\@@href}%
	\providecommand \@@href[1]{\endgroup#1\@@endlink}%
	\providecommand \@sanitize@url [0]{\catcode `\\12\catcode `\$12\catcode
	  `\&12\catcode `\#12\catcode `\^12\catcode `\_12\catcode `\%12\relax}%
	\providecommand \@@startlink[1]{}%
	\providecommand \@@endlink[0]{}%
	\providecommand \url  [0]{\begingroup\@sanitize@url \@url }%
	\providecommand \@url [1]{\endgroup\@href {#1}{\urlprefix }}%
	\providecommand \urlprefix  [0]{URL }%
	\providecommand \Eprint [0]{\href }%
	\providecommand \doibase [0]{https://doi.org/}%
	\providecommand \selectlanguage [0]{\@gobble}%
	\providecommand \bibinfo  [0]{\@secondoftwo}%
	\providecommand \bibfield  [0]{\@secondoftwo}%
	\providecommand \translation [1]{[#1]}%
	\providecommand \BibitemOpen [0]{}%
	\providecommand \bibitemStop [0]{}%
	\providecommand \bibitemNoStop [0]{.\EOS\space}%
	\providecommand \EOS [0]{\spacefactor3000\relax}%
	\providecommand \BibitemShut  [1]{\csname bibitem#1\endcsname}%
	\let\auto@bib@innerbib\@empty
	\bibitem [{\citenamefont {Behler}\ and\ \citenamefont
	  {Parrinello}(2007)}]{behler_2007}%
	  \BibitemOpen
	  \bibfield  {author} {\bibinfo {author} {\bibfnamefont {J.}~\bibnamefont
	  {Behler}}\ and\ \bibinfo {author} {\bibfnamefont {M.}~\bibnamefont
	  {Parrinello}},\ }\bibfield  {title} {\bibinfo {title} {Generalized
	  neural-network representation of high-dimensional potential-energy
	  surfaces},\ }\href {https://doi.org/10.1103/PhysRevLett.98.146401} {\bibfield
	   {journal} {\bibinfo  {journal} {Phys. Rev. Lett.}\ }\textbf {\bibinfo
	  {volume} {98}},\ \bibinfo {pages} {146401} (\bibinfo {year}
	  {2007})}\BibitemShut {NoStop}%
	\bibitem [{\citenamefont {Schütt}\ \emph {et~al.}(2018)\citenamefont
	  {Schütt}, \citenamefont {Sauceda}, \citenamefont {Kindermans}, \citenamefont
	  {Tkatchenko},\ and\ \citenamefont {Müller}}]{schutt_schnet_2017}%
	  \BibitemOpen
	  \bibfield  {author} {\bibinfo {author} {\bibfnamefont {K.~T.}\ \bibnamefont
	  {Schütt}}, \bibinfo {author} {\bibfnamefont {H.~E.}\ \bibnamefont
	  {Sauceda}}, \bibinfo {author} {\bibfnamefont {P.-J.}\ \bibnamefont
	  {Kindermans}}, \bibinfo {author} {\bibfnamefont {A.}~\bibnamefont
	  {Tkatchenko}},\ and\ \bibinfo {author} {\bibfnamefont {K.-R.}\ \bibnamefont
	  {Müller}},\ }\bibfield  {title} {\bibinfo {title} {{SchNet} – {A} deep
	  learning architecture for molecules and materials},\ }\href
	  {https://doi.org/10.1063/1.5019779} {\bibfield  {journal} {\bibinfo
	  {journal} {J. Chem. Phys.}\ }\textbf {\bibinfo {volume} {148}},\ \bibinfo
	  {pages} {241722} (\bibinfo {year} {2018})}\BibitemShut {NoStop}%
	\bibitem [{\citenamefont {Zhang}\ \emph {et~al.}(2018)\citenamefont {Zhang},
	  \citenamefont {Han}, \citenamefont {Wang}, \citenamefont {Car},\ and\
	  \citenamefont {E}}]{zhang_deep_2018}%
	  \BibitemOpen
	  \bibfield  {author} {\bibinfo {author} {\bibfnamefont {L.}~\bibnamefont
	  {Zhang}}, \bibinfo {author} {\bibfnamefont {J.}~\bibnamefont {Han}}, \bibinfo
	  {author} {\bibfnamefont {H.}~\bibnamefont {Wang}}, \bibinfo {author}
	  {\bibfnamefont {R.}~\bibnamefont {Car}},\ and\ \bibinfo {author}
	  {\bibfnamefont {W.}~\bibnamefont {E}},\ }\bibfield  {title} {\bibinfo {title}
	  {Deep {Potential} {Molecular} {Dynamics}: {A} {Scalable} {Model} with the
	  {Accuracy} of {Quantum} {Mechanics}},\ }\href
	  {https://doi.org/10.1103/PhysRevLett.120.143001} {\bibfield  {journal}
	  {\bibinfo  {journal} {Phys. Rev. Lett.}\ }\textbf {\bibinfo {volume} {120}},\
	  \bibinfo {pages} {143001} (\bibinfo {year} {2018})}\BibitemShut {NoStop}%
	\bibitem [{\citenamefont {Unke}\ \emph {et~al.}(2021)\citenamefont {Unke},
	  \citenamefont {Bogojeski}, \citenamefont {Gastegger}, \citenamefont {Geiger},
	  \citenamefont {Smidt},\ and\ \citenamefont {M\"{u}ller}}]{Unke2021}%
	  \BibitemOpen
	  \bibfield  {author} {\bibinfo {author} {\bibfnamefont {O.~T.}\ \bibnamefont
	  {Unke}}, \bibinfo {author} {\bibfnamefont {M.}~\bibnamefont {Bogojeski}},
	  \bibinfo {author} {\bibfnamefont {M.}~\bibnamefont {Gastegger}}, \bibinfo
	  {author} {\bibfnamefont {M.}~\bibnamefont {Geiger}}, \bibinfo {author}
	  {\bibfnamefont {T.}~\bibnamefont {Smidt}},\ and\ \bibinfo {author}
	  {\bibfnamefont {K.-R.}\ \bibnamefont {M\"{u}ller}},\ }\bibfield  {title}
	  {\bibinfo {title} {{SE}(3)-equivariant prediction of molecular wavefunctions
	  and electronic densities},\ }in\ \href
	  {https://openreview.net/forum?id=auGY2UQfhSu} {\emph {\bibinfo {booktitle}
	  {Advances in Neural Information Processing Systems}}}\ (\bibinfo  {publisher}
	  {Curran Associates, Inc.},\ \bibinfo {year} {2021})\ p.\ \bibinfo {pages}
	  {14434}\BibitemShut {NoStop}%
	\bibitem [{\citenamefont {Gu}\ \emph {et~al.}(2022)\citenamefont {Gu},
	  \citenamefont {Zhang},\ and\ \citenamefont {Feng}}]{Gu2022}%
	  \BibitemOpen
	  \bibfield  {author} {\bibinfo {author} {\bibfnamefont {Q.}~\bibnamefont
	  {Gu}}, \bibinfo {author} {\bibfnamefont {L.}~\bibnamefont {Zhang}},\ and\
	  \bibinfo {author} {\bibfnamefont {J.}~\bibnamefont {Feng}},\ }\bibfield
	  {title} {\bibinfo {title} {Neural network representation of electronic
	  structure from ab initio molecular dynamics},\ }\href
	  {https://doi.org/https://doi.org/10.1016/j.scib.2021.09.010} {\bibfield
	  {journal} {\bibinfo  {journal} {Sci. Bull.}\ }\textbf {\bibinfo {volume}
	  {67}},\ \bibinfo {pages} {29} (\bibinfo {year} {2022})}\BibitemShut {NoStop}%
	\bibitem [{\citenamefont {Batzner}\ \emph {et~al.}(2022)\citenamefont
	  {Batzner}, \citenamefont {Musaelian}, \citenamefont {Sun}, \citenamefont
	  {Geiger}, \citenamefont {Mailoa}, \citenamefont {Kornbluth}, \citenamefont
	  {Molinari}, \citenamefont {Smidt},\ and\ \citenamefont
	  {Kozinsky}}]{batzner_e3-equivariant_2022}%
	  \BibitemOpen
	  \bibfield  {author} {\bibinfo {author} {\bibfnamefont {S.}~\bibnamefont
	  {Batzner}}, \bibinfo {author} {\bibfnamefont {A.}~\bibnamefont {Musaelian}},
	  \bibinfo {author} {\bibfnamefont {L.}~\bibnamefont {Sun}}, \bibinfo {author}
	  {\bibfnamefont {M.}~\bibnamefont {Geiger}}, \bibinfo {author} {\bibfnamefont
	  {J.~P.}\ \bibnamefont {Mailoa}}, \bibinfo {author} {\bibfnamefont
	  {M.}~\bibnamefont {Kornbluth}}, \bibinfo {author} {\bibfnamefont
	  {N.}~\bibnamefont {Molinari}}, \bibinfo {author} {\bibfnamefont {T.~E.}\
	  \bibnamefont {Smidt}},\ and\ \bibinfo {author} {\bibfnamefont
	  {B.}~\bibnamefont {Kozinsky}},\ }\bibfield  {title} {\bibinfo {title}
	  {E(3)-equivariant graph neural networks for data-efficient and accurate
	  interatomic potentials},\ }\href {https://doi.org/10.1038/s41467-022-29939-5}
	  {\bibfield  {journal} {\bibinfo  {journal} {Nat. Commun.}\ }\textbf {\bibinfo
	  {volume} {13}},\ \bibinfo {pages} {2453} (\bibinfo {year}
	  {2022})}\BibitemShut {NoStop}%
	\bibitem [{\citenamefont {Li}\ \emph {et~al.}(2022)\citenamefont {Li},
	  \citenamefont {Wang}, \citenamefont {Zou}, \citenamefont {Ye}, \citenamefont
	  {Xu}, \citenamefont {Gong}, \citenamefont {Duan},\ and\ \citenamefont
	  {Xu}}]{li2022deep}%
	  \BibitemOpen
	  \bibfield  {author} {\bibinfo {author} {\bibfnamefont {H.}~\bibnamefont
	  {Li}}, \bibinfo {author} {\bibfnamefont {Z.}~\bibnamefont {Wang}}, \bibinfo
	  {author} {\bibfnamefont {N.}~\bibnamefont {Zou}}, \bibinfo {author}
	  {\bibfnamefont {M.}~\bibnamefont {Ye}}, \bibinfo {author} {\bibfnamefont
	  {R.}~\bibnamefont {Xu}}, \bibinfo {author} {\bibfnamefont {X.}~\bibnamefont
	  {Gong}}, \bibinfo {author} {\bibfnamefont {W.}~\bibnamefont {Duan}},\ and\
	  \bibinfo {author} {\bibfnamefont {Y.}~\bibnamefont {Xu}},\ }\bibfield
	  {title} {\bibinfo {title} {Deep-learning density functional theory
	  hamiltonian for efficient ab initio electronic-structure calculation},\
	  }\href {https://doi.org/10.1038/s43588-022-00265-6} {\bibfield  {journal}
	  {\bibinfo  {journal} {Nat. Comput. Sci.}\ }\textbf {\bibinfo {volume} {2}},\
	  \bibinfo {pages} {367} (\bibinfo {year} {2022})}\BibitemShut {NoStop}%
	\bibitem [{\citenamefont {Gong}\ \emph {et~al.}(2023)\citenamefont {Gong},
	  \citenamefont {Li}, \citenamefont {Zou}, \citenamefont {Xu}, \citenamefont
	  {Duan},\ and\ \citenamefont {Xu}}]{gong_general_2023}%
	  \BibitemOpen
	  \bibfield  {author} {\bibinfo {author} {\bibfnamefont {X.}~\bibnamefont
	  {Gong}}, \bibinfo {author} {\bibfnamefont {H.}~\bibnamefont {Li}}, \bibinfo
	  {author} {\bibfnamefont {N.}~\bibnamefont {Zou}}, \bibinfo {author}
	  {\bibfnamefont {R.}~\bibnamefont {Xu}}, \bibinfo {author} {\bibfnamefont
	  {W.}~\bibnamefont {Duan}},\ and\ \bibinfo {author} {\bibfnamefont
	  {Y.}~\bibnamefont {Xu}},\ }\bibfield  {title} {\bibinfo {title} {General
	  framework for {E}(3)-equivariant neural network representation of density
	  functional theory {Hamiltonian}},\ }\href {http://arxiv.org/abs/2210.13955}
	  {\bibfield  {journal} {\bibinfo  {journal} {Nat. Commun.}\ }\textbf {\bibinfo
	  {volume} {14}},\ \bibinfo {pages} {2848} (\bibinfo {year}
	  {2023})}\BibitemShut {NoStop}%
	\bibitem [{\citenamefont {Tang}\ \emph {et~al.}(2023)\citenamefont {Tang},
	  \citenamefont {Li}, \citenamefont {Lin}, \citenamefont {Gong}, \citenamefont
	  {Jin}, \citenamefont {He}, \citenamefont {Jiang}, \citenamefont {Ren},
	  \citenamefont {Duan},\ and\ \citenamefont {Xu}}]{tang2023efficient}%
	  \BibitemOpen
	  \bibfield  {author} {\bibinfo {author} {\bibfnamefont {Z.}~\bibnamefont
	  {Tang}}, \bibinfo {author} {\bibfnamefont {H.}~\bibnamefont {Li}}, \bibinfo
	  {author} {\bibfnamefont {P.}~\bibnamefont {Lin}}, \bibinfo {author}
	  {\bibfnamefont {X.}~\bibnamefont {Gong}}, \bibinfo {author} {\bibfnamefont
	  {G.}~\bibnamefont {Jin}}, \bibinfo {author} {\bibfnamefont {L.}~\bibnamefont
	  {He}}, \bibinfo {author} {\bibfnamefont {H.}~\bibnamefont {Jiang}}, \bibinfo
	  {author} {\bibfnamefont {X.}~\bibnamefont {Ren}}, \bibinfo {author}
	  {\bibfnamefont {W.}~\bibnamefont {Duan}},\ and\ \bibinfo {author}
	  {\bibfnamefont {Y.}~\bibnamefont {Xu}},\ }\bibfield  {title} {\bibinfo
	  {title} {Efficient hybrid density functional calculation by deep learning},\
	  }\href {https://arxiv.org/abs/2302.08221} {\bibfield  {journal} {\bibinfo
	  {journal} {arXiv:2302.08221}\ } (\bibinfo {year} {2023})}\BibitemShut
	  {NoStop}%
	\bibitem [{\citenamefont {Li}\ \emph {et~al.}(2023)\citenamefont {Li},
	  \citenamefont {Tang}, \citenamefont {Gong}, \citenamefont {Zou},
	  \citenamefont {Duan},\ and\ \citenamefont {Xu}}]{li2023deep}%
	  \BibitemOpen
	  \bibfield  {author} {\bibinfo {author} {\bibfnamefont {H.}~\bibnamefont
	  {Li}}, \bibinfo {author} {\bibfnamefont {Z.}~\bibnamefont {Tang}}, \bibinfo
	  {author} {\bibfnamefont {X.}~\bibnamefont {Gong}}, \bibinfo {author}
	  {\bibfnamefont {N.}~\bibnamefont {Zou}}, \bibinfo {author} {\bibfnamefont
	  {W.}~\bibnamefont {Duan}},\ and\ \bibinfo {author} {\bibfnamefont
	  {Y.}~\bibnamefont {Xu}},\ }\bibfield  {title} {\bibinfo {title}
	  {Deep-learning electronic-structure calculation of magnetic
	  superstructures},\ }\href
	  {https://www.nature.com/articles/s43588-023-00424-3} {\bibfield  {journal}
	  {\bibinfo  {journal} {Nat. Comput. Sci.}\ }\textbf {\bibinfo {volume} {3}},\
	  \bibinfo {pages} {321} (\bibinfo {year} {2023})}\BibitemShut {NoStop}%
	\bibitem [{\citenamefont {Li}\ \emph {et~al.}(2024{\natexlab{a}})\citenamefont
	  {Li}, \citenamefont {Tang}, \citenamefont {Fu}, \citenamefont {Dong},
	  \citenamefont {Zou}, \citenamefont {Gong}, \citenamefont {Duan},\ and\
	  \citenamefont {Xu}}]{li2024deep}%
	  \BibitemOpen
	  \bibfield  {author} {\bibinfo {author} {\bibfnamefont {H.}~\bibnamefont
	  {Li}}, \bibinfo {author} {\bibfnamefont {Z.}~\bibnamefont {Tang}}, \bibinfo
	  {author} {\bibfnamefont {J.}~\bibnamefont {Fu}}, \bibinfo {author}
	  {\bibfnamefont {W.-H.}\ \bibnamefont {Dong}}, \bibinfo {author}
	  {\bibfnamefont {N.}~\bibnamefont {Zou}}, \bibinfo {author} {\bibfnamefont
	  {X.}~\bibnamefont {Gong}}, \bibinfo {author} {\bibfnamefont {W.}~\bibnamefont
	  {Duan}},\ and\ \bibinfo {author} {\bibfnamefont {Y.}~\bibnamefont {Xu}},\
	  }\bibfield  {title} {\bibinfo {title} {Deep-learning density functional
	  perturbation theory},\ }\href
	  {https://doi.org/10.1103/PhysRevLett.132.096401} {\bibfield  {journal}
	  {\bibinfo  {journal} {Phys. Rev. Lett.}\ }\textbf {\bibinfo {volume} {132}},\
	  \bibinfo {pages} {096401} (\bibinfo {year} {2024}{\natexlab{a}})}\BibitemShut
	  {NoStop}%
	\bibitem [{\citenamefont {Wang}\ \emph
	  {et~al.}(2024{\natexlab{a}})\citenamefont {Wang}, \citenamefont {Li},
	  \citenamefont {Tang}, \citenamefont {Tao}, \citenamefont {Wang},
	  \citenamefont {Yuan}, \citenamefont {Chen}, \citenamefont {Duan},\ and\
	  \citenamefont {Xu}}]{wang2024deeph}%
	  \BibitemOpen
	  \bibfield  {author} {\bibinfo {author} {\bibfnamefont {Y.}~\bibnamefont
	  {Wang}}, \bibinfo {author} {\bibfnamefont {H.}~\bibnamefont {Li}}, \bibinfo
	  {author} {\bibfnamefont {Z.}~\bibnamefont {Tang}}, \bibinfo {author}
	  {\bibfnamefont {H.}~\bibnamefont {Tao}}, \bibinfo {author} {\bibfnamefont
	  {Y.}~\bibnamefont {Wang}}, \bibinfo {author} {\bibfnamefont {Z.}~\bibnamefont
	  {Yuan}}, \bibinfo {author} {\bibfnamefont {Z.}~\bibnamefont {Chen}}, \bibinfo
	  {author} {\bibfnamefont {W.}~\bibnamefont {Duan}},\ and\ \bibinfo {author}
	  {\bibfnamefont {Y.}~\bibnamefont {Xu}},\ }\bibfield  {title} {\bibinfo
	  {title} {{DeepH-2}: Enhancing deep-learning electronic structure via an
	  equivariant local-coordinate transformer},\ }\href
	  {https://doi.org/10.48550/arXiv.2401.17015} {\bibfield  {journal} {\bibinfo
	  {journal} {arXiv:2401.17015,}\ } (\bibinfo {year}
	  {2024}{\natexlab{a}})}\BibitemShut {NoStop}%
	\bibitem [{\citenamefont {Wang}\ \emph
	  {et~al.}(2024{\natexlab{b}})\citenamefont {Wang}, \citenamefont {Li},
	  \citenamefont {Tang}, \citenamefont {Li}, \citenamefont {Yuan}, \citenamefont
	  {Tao}, \citenamefont {Zou}, \citenamefont {Bao}, \citenamefont {Liang},
	  \citenamefont {Chen}, \citenamefont {Xu}, \citenamefont {Bian}, \citenamefont
	  {Xu}, \citenamefont {Wang}, \citenamefont {Si}, \citenamefont {Duan},\ and\
	  \citenamefont {Xu}}]{WANG2024}%
	  \BibitemOpen
	  \bibfield  {author} {\bibinfo {author} {\bibfnamefont {Y.}~\bibnamefont
	  {Wang}}, \bibinfo {author} {\bibfnamefont {Y.}~\bibnamefont {Li}}, \bibinfo
	  {author} {\bibfnamefont {Z.}~\bibnamefont {Tang}}, \bibinfo {author}
	  {\bibfnamefont {H.}~\bibnamefont {Li}}, \bibinfo {author} {\bibfnamefont
	  {Z.}~\bibnamefont {Yuan}}, \bibinfo {author} {\bibfnamefont {H.}~\bibnamefont
	  {Tao}}, \bibinfo {author} {\bibfnamefont {N.}~\bibnamefont {Zou}}, \bibinfo
	  {author} {\bibfnamefont {T.}~\bibnamefont {Bao}}, \bibinfo {author}
	  {\bibfnamefont {X.}~\bibnamefont {Liang}}, \bibinfo {author} {\bibfnamefont
	  {Z.}~\bibnamefont {Chen}}, \bibinfo {author} {\bibfnamefont {S.}~\bibnamefont
	  {Xu}}, \bibinfo {author} {\bibfnamefont {C.}~\bibnamefont {Bian}}, \bibinfo
	  {author} {\bibfnamefont {Z.}~\bibnamefont {Xu}}, \bibinfo {author}
	  {\bibfnamefont {C.}~\bibnamefont {Wang}}, \bibinfo {author} {\bibfnamefont
	  {C.}~\bibnamefont {Si}}, \bibinfo {author} {\bibfnamefont {W.}~\bibnamefont
	  {Duan}},\ and\ \bibinfo {author} {\bibfnamefont {Y.}~\bibnamefont {Xu}},\
	  }\bibfield  {title} {\bibinfo {title} {Universal materials model of
	  deep-learning density functional theory hamiltonian},\ }\href
	  {https://www.sciencedirect.com/science/article/pii/S2095927324004079}
	  {\bibfield  {journal} {\bibinfo  {journal} {Sci. Bull.}\ } (\bibinfo {year}
	  {2024}{\natexlab{b}})}\BibitemShut {NoStop}%
	\bibitem [{\citenamefont {Li}\ \emph {et~al.}(2024{\natexlab{b}})\citenamefont
	  {Li}, \citenamefont {Tang}, \citenamefont {Chen}, \citenamefont {Sun},
	  \citenamefont {Zhao}, \citenamefont {Li}, \citenamefont {Tao}, \citenamefont
	  {Yuan}, \citenamefont {Duan},\ and\ \citenamefont {Xu}}]{li2024neural}%
	  \BibitemOpen
	  \bibfield  {author} {\bibinfo {author} {\bibfnamefont {Y.}~\bibnamefont
	  {Li}}, \bibinfo {author} {\bibfnamefont {Z.}~\bibnamefont {Tang}}, \bibinfo
	  {author} {\bibfnamefont {Z.}~\bibnamefont {Chen}}, \bibinfo {author}
	  {\bibfnamefont {M.}~\bibnamefont {Sun}}, \bibinfo {author} {\bibfnamefont
	  {B.}~\bibnamefont {Zhao}}, \bibinfo {author} {\bibfnamefont {H.}~\bibnamefont
	  {Li}}, \bibinfo {author} {\bibfnamefont {H.}~\bibnamefont {Tao}}, \bibinfo
	  {author} {\bibfnamefont {Z.}~\bibnamefont {Yuan}}, \bibinfo {author}
	  {\bibfnamefont {W.}~\bibnamefont {Duan}},\ and\ \bibinfo {author}
	  {\bibfnamefont {Y.}~\bibnamefont {Xu}},\ }\bibfield  {title} {\bibinfo
	  {title} {Neural-network density functional theory},\ }\href
	  {https://arxiv.org/abs/2403.11287} {\bibfield  {journal} {\bibinfo  {journal}
	  {arXiv:2403.11287}\ } (\bibinfo {year} {2024}{\natexlab{b}})}\BibitemShut
	  {NoStop}%
	\bibitem [{\citenamefont {Yuan}\ \emph {et~al.}(2024)\citenamefont {Yuan},
	  \citenamefont {Xu}, \citenamefont {Li}, \citenamefont {Cheng}, \citenamefont
	  {Tao}, \citenamefont {Tang}, \citenamefont {Zhou}, \citenamefont {Duan},\
	  and\ \citenamefont {Xu}}]{yuan_equivariant_2024}%
	  \BibitemOpen
	  \bibfield  {author} {\bibinfo {author} {\bibfnamefont {Z.}~\bibnamefont
	  {Yuan}}, \bibinfo {author} {\bibfnamefont {Z.}~\bibnamefont {Xu}}, \bibinfo
	  {author} {\bibfnamefont {H.}~\bibnamefont {Li}}, \bibinfo {author}
	  {\bibfnamefont {X.}~\bibnamefont {Cheng}}, \bibinfo {author} {\bibfnamefont
	  {H.}~\bibnamefont {Tao}}, \bibinfo {author} {\bibfnamefont {Z.}~\bibnamefont
	  {Tang}}, \bibinfo {author} {\bibfnamefont {Z.}~\bibnamefont {Zhou}}, \bibinfo
	  {author} {\bibfnamefont {W.}~\bibnamefont {Duan}},\ and\ \bibinfo {author}
	  {\bibfnamefont {Y.}~\bibnamefont {Xu}},\ }\bibfield  {title} {\bibinfo
	  {title} {Equivariant neural network force fields for magnetic materials},\
	  }\href {https://doi.org/10.1007/s44214-024-00055-3} {\bibfield  {journal}
	  {\bibinfo  {journal} {Quantum. Front.}\ }\textbf {\bibinfo {volume} {3}},\
	  \bibinfo {pages} {8} (\bibinfo {year} {2024})}\BibitemShut {NoStop}%
	\bibitem [{\citenamefont {Tang}\ \emph {et~al.}(2024)\citenamefont {Tang},
	  \citenamefont {Zou}, \citenamefont {Li}, \citenamefont {Wang}, \citenamefont
	  {Yuan}, \citenamefont {Tao}, \citenamefont {Li}, \citenamefont {Chen},
	  \citenamefont {Zhao}, \citenamefont {Sun}, \citenamefont {Jiang},
	  \citenamefont {Duan},\ and\ \citenamefont {Xu}}]{tang2024densitymatrix}%
	  \BibitemOpen
	  \bibfield  {author} {\bibinfo {author} {\bibfnamefont {Z.}~\bibnamefont
	  {Tang}}, \bibinfo {author} {\bibfnamefont {N.}~\bibnamefont {Zou}}, \bibinfo
	  {author} {\bibfnamefont {H.}~\bibnamefont {Li}}, \bibinfo {author}
	  {\bibfnamefont {Y.}~\bibnamefont {Wang}}, \bibinfo {author} {\bibfnamefont
	  {Z.}~\bibnamefont {Yuan}}, \bibinfo {author} {\bibfnamefont {H.}~\bibnamefont
	  {Tao}}, \bibinfo {author} {\bibfnamefont {Y.}~\bibnamefont {Li}}, \bibinfo
	  {author} {\bibfnamefont {Z.}~\bibnamefont {Chen}}, \bibinfo {author}
	  {\bibfnamefont {B.}~\bibnamefont {Zhao}}, \bibinfo {author} {\bibfnamefont
	  {M.}~\bibnamefont {Sun}}, \bibinfo {author} {\bibfnamefont {H.}~\bibnamefont
	  {Jiang}}, \bibinfo {author} {\bibfnamefont {W.}~\bibnamefont {Duan}},\ and\
	  \bibinfo {author} {\bibfnamefont {Y.}~\bibnamefont {Xu}},\ }\bibfield
	  {title} {\bibinfo {title} {Improving density matrix electronic structure
	  method by deep learning},\ }\href {https://arxiv.org/abs/2406.17561}
	  {\bibfield  {journal} {\bibinfo  {journal} {arXiv:2406.17561}\ } (\bibinfo
	  {year} {2024})}\BibitemShut {NoStop}%
	\bibitem [{\citenamefont {Yu}\ \emph {et~al.}(2023)\citenamefont {Yu},
	  \citenamefont {Xu}, \citenamefont {Qian}, \citenamefont {Qian},\ and\
	  \citenamefont {Ji}}]{yu2023efficient}%
	  \BibitemOpen
	  \bibfield  {author} {\bibinfo {author} {\bibfnamefont {H.}~\bibnamefont
	  {Yu}}, \bibinfo {author} {\bibfnamefont {Z.}~\bibnamefont {Xu}}, \bibinfo
	  {author} {\bibfnamefont {X.}~\bibnamefont {Qian}}, \bibinfo {author}
	  {\bibfnamefont {X.}~\bibnamefont {Qian}},\ and\ \bibinfo {author}
	  {\bibfnamefont {S.}~\bibnamefont {Ji}},\ }\bibfield  {title} {\bibinfo
	  {title} {Efficient and equivariant graph networks for predicting quantum
	  {Hamiltonian}},\ }in\ \href {https://proceedings.mlr.press/v202/yu23i.html}
	  {\emph {\bibinfo {booktitle} {International Conference on Machine
	  Learning}}}\ (\bibinfo {organization} {PMLR},\ \bibinfo {year} {2023})\ pp.\
	  \bibinfo {pages} {40412--40424}\BibitemShut {NoStop}%
	\bibitem [{\citenamefont {Sch{\"u}tt}\ \emph {et~al.}(2021)\citenamefont
	  {Sch{\"u}tt}, \citenamefont {Unke},\ and\ \citenamefont
	  {Gastegger}}]{phisnet_2021}%
	  \BibitemOpen
	  \bibfield  {author} {\bibinfo {author} {\bibfnamefont {K.}~\bibnamefont
	  {Sch{\"u}tt}}, \bibinfo {author} {\bibfnamefont {O.}~\bibnamefont {Unke}},\
	  and\ \bibinfo {author} {\bibfnamefont {M.}~\bibnamefont {Gastegger}},\
	  }\bibfield  {title} {\bibinfo {title} {Equivariant message passing for the
	  prediction of tensorial properties and molecular spectra},\ }in\ \href
	  {https://proceedings.mlr.press/v139/schutt21a.html} {\emph {\bibinfo
	  {booktitle} {Proceedings of the 38th International Conference on Machine
	  Learning}}},\ Vol.\ \bibinfo {volume} {139}\ (\bibinfo {year} {2021})\ pp.\
	  \bibinfo {pages} {9377--9388}\BibitemShut {NoStop}%
	\bibitem [{\citenamefont {Liao}\ \emph {et~al.}(2024)\citenamefont {Liao},
	  \citenamefont {Wood}, \citenamefont {Das},\ and\ \citenamefont
	  {Smidt}}]{liao2023equiformerv2}%
	  \BibitemOpen
	  \bibfield  {author} {\bibinfo {author} {\bibfnamefont {Y.-L.}\ \bibnamefont
	  {Liao}}, \bibinfo {author} {\bibfnamefont {B.~M.}\ \bibnamefont {Wood}},
	  \bibinfo {author} {\bibfnamefont {A.}~\bibnamefont {Das}},\ and\ \bibinfo
	  {author} {\bibfnamefont {T.}~\bibnamefont {Smidt}},\ }\bibfield  {title}
	  {\bibinfo {title} {Equiformerv2: Improved equivariant transformer for scaling
	  to higher-degree representations},\ }in\ \href
	  {https://openreview.net/forum?id=mCOBKZmrzD} {\emph {\bibinfo {booktitle}
	  {The Twelfth International Conference on Learning Representations}}}\
	  (\bibinfo {year} {2024})\BibitemShut {NoStop}%
	\bibitem [{\citenamefont {Kohn}(1996)}]{Kohn1996}%
	  \BibitemOpen
	  \bibfield  {author} {\bibinfo {author} {\bibfnamefont {W.}~\bibnamefont
	  {Kohn}},\ }\bibfield  {title} {\bibinfo {title} {Density functional and
	  density matrix method scaling linearly with the number of atoms},\ }\href
	  {https://doi.org/10.1103/PhysRevLett.76.3168} {\bibfield  {journal} {\bibinfo
	   {journal} {Phys. Rev. Lett.}\ }\textbf {\bibinfo {volume} {76}},\ \bibinfo
	  {pages} {3168} (\bibinfo {year} {1996})}\BibitemShut {NoStop}%
	\bibitem [{\citenamefont {Prodan}\ and\ \citenamefont
	  {Kohn}(2005)}]{Prodan2005}%
	  \BibitemOpen
	  \bibfield  {author} {\bibinfo {author} {\bibfnamefont {E.}~\bibnamefont
	  {Prodan}}\ and\ \bibinfo {author} {\bibfnamefont {W.}~\bibnamefont {Kohn}},\
	  }\bibfield  {title} {\bibinfo {title} {Nearsightedness of electronic
	  matter},\ }\href {https://doi.org/10.1073/pnas.0505436102} {\bibfield
	  {journal} {\bibinfo  {journal} {Proc. Natl. Acad. Sci. U.S.A.}\ }\textbf
	  {\bibinfo {volume} {102}},\ \bibinfo {pages} {11635} (\bibinfo {year}
	  {2005})}\BibitemShut {NoStop}%
	\bibitem [{\citenamefont {Martin}(2004)}]{Martin2004}%
	  \BibitemOpen
	  \bibfield  {author} {\bibinfo {author} {\bibfnamefont {R.~M.}\ \bibnamefont
	  {Martin}},\ }\href {https://doi.org/10.1017/CBO9780511805769} {\emph
	  {\bibinfo {title} {Electronic Structure: Basic Theory and Practical
	  Methods}}}\ (\bibinfo  {publisher} {Cambridge University Press, Cambridge,
	  England},\ \bibinfo {year} {2004})\BibitemShut {NoStop}%
	\bibitem [{\citenamefont {Soler}\ \emph {et~al.}(2002)\citenamefont {Soler},
	  \citenamefont {Artacho}, \citenamefont {Gale}, \citenamefont {Garc{\'\i}a},
	  \citenamefont {Junquera}, \citenamefont {Ordej{\'o}n},\ and\ \citenamefont
	  {S{\'a}nchez-Portal}}]{soler2002siesta}%
	  \BibitemOpen
	  \bibfield  {author} {\bibinfo {author} {\bibfnamefont {J.~M.}\ \bibnamefont
	  {Soler}}, \bibinfo {author} {\bibfnamefont {E.}~\bibnamefont {Artacho}},
	  \bibinfo {author} {\bibfnamefont {J.~D.}\ \bibnamefont {Gale}}, \bibinfo
	  {author} {\bibfnamefont {A.}~\bibnamefont {Garc{\'\i}a}}, \bibinfo {author}
	  {\bibfnamefont {J.}~\bibnamefont {Junquera}}, \bibinfo {author}
	  {\bibfnamefont {P.}~\bibnamefont {Ordej{\'o}n}},\ and\ \bibinfo {author}
	  {\bibfnamefont {D.}~\bibnamefont {S{\'a}nchez-Portal}},\ }\bibfield  {title}
	  {\bibinfo {title} {The {SIESTA} method for ab initio order-{N} materials
	  simulation},\ }\href
	  {https://iopscience.iop.org/article/10.1088/0953-8984/14/11/302} {\bibfield
	  {journal} {\bibinfo  {journal} {J. Phys.: Condens. Matter}\ }\textbf
	  {\bibinfo {volume} {14}},\ \bibinfo {pages} {2745} (\bibinfo {year}
	  {2002})}\BibitemShut {NoStop}%
	\bibitem [{\citenamefont {Hamann}\ \emph {et~al.}(1979)\citenamefont {Hamann},
	  \citenamefont {Schl{\"u}ter},\ and\ \citenamefont {Chiang}}]{ncpp1979}%
	  \BibitemOpen
	  \bibfield  {author} {\bibinfo {author} {\bibfnamefont {D.}~\bibnamefont
	  {Hamann}}, \bibinfo {author} {\bibfnamefont {M.}~\bibnamefont
	  {Schl{\"u}ter}},\ and\ \bibinfo {author} {\bibfnamefont {C.}~\bibnamefont
	  {Chiang}},\ }\bibfield  {title} {\bibinfo {title} {Norm-conserving
	  pseudopotentials},\ }\href {https://doi.org/10.1103/PhysRevLett.43.1494}
	  {\bibfield  {journal} {\bibinfo  {journal} {Phys. Rev. Lett.}\ }\textbf
	  {\bibinfo {volume} {43}},\ \bibinfo {pages} {1494} (\bibinfo {year}
	  {1979})}\BibitemShut {NoStop}%
	\bibitem [{\citenamefont {Kleinman}\ and\ \citenamefont
	  {Bylander}(1982)}]{kbproj1982}%
	  \BibitemOpen
	  \bibfield  {author} {\bibinfo {author} {\bibfnamefont {L.}~\bibnamefont
	  {Kleinman}}\ and\ \bibinfo {author} {\bibfnamefont {D.}~\bibnamefont
	  {Bylander}},\ }\bibfield  {title} {\bibinfo {title} {Efficacious form for
	  model pseudopotentials},\ }\href
	  {https://doi.org/10.1103/PhysRevLett.48.1425} {\bibfield  {journal} {\bibinfo
	   {journal} {Phys. Rev. Lett.}\ }\textbf {\bibinfo {volume} {48}},\ \bibinfo
	  {pages} {1425} (\bibinfo {year} {1982})}\BibitemShut {NoStop}%
	\bibitem [{sup()}]{supp}%
	  \BibitemOpen
	  \href@noop {} {\ }\bibinfo {note} {See Supplementary Material for
	  details}\BibitemShut {NoStop}%
	\bibitem [{\citenamefont {Passaro}\ and\ \citenamefont
	  {Zitnick}(2023)}]{passaro2023reducing}%
	  \BibitemOpen
	  \bibfield  {author} {\bibinfo {author} {\bibfnamefont {S.}~\bibnamefont
	  {Passaro}}\ and\ \bibinfo {author} {\bibfnamefont {C.~L.}\ \bibnamefont
	  {Zitnick}},\ }\bibfield  {title} {\bibinfo {title} {Reducing {SO}(3)
	  convolutions to {SO}(2) for efficient equivariant {GNN}s},\ }in\ \href
	  {https://dl.acm.org/doi/10.5555/3618408.3619548} {\emph {\bibinfo {booktitle}
	  {Proceedings of the 40th International Conference on Machine Learning}}}\
	  (\bibinfo {year} {2023})\BibitemShut {NoStop}%
	\bibitem [{\citenamefont {Hoffmann}\ \emph {et~al.}(2016)\citenamefont
	  {Hoffmann}, \citenamefont {Kabanov}, \citenamefont {Golov},\ and\
	  \citenamefont {Proserpio}}]{SACADA2016}%
	  \BibitemOpen
	  \bibfield  {author} {\bibinfo {author} {\bibfnamefont {R.}~\bibnamefont
	  {Hoffmann}}, \bibinfo {author} {\bibfnamefont {A.~A.}\ \bibnamefont
	  {Kabanov}}, \bibinfo {author} {\bibfnamefont {A.~A.}\ \bibnamefont {Golov}},\
	  and\ \bibinfo {author} {\bibfnamefont {D.~M.}\ \bibnamefont {Proserpio}},\
	  }\bibfield  {title} {\bibinfo {title} {Homo citans and carbon allotropes: For
	  an ethics of citation},\ }\href {https://doi.org/10.1002/anie.201600655}
	  {\bibfield  {journal} {\bibinfo  {journal} {Angew. Chem. Int. Ed.}\ }\textbf
	  {\bibinfo {volume} {55}},\ \bibinfo {pages} {10962} (\bibinfo {year}
	  {2016})}\BibitemShut {NoStop}%
	\end{thebibliography}
\end{document}